\documentclass[preprint,showpacs,preprintnumbers,amsmath,amssymb]{revtex4}

\usepackage[dvips]{graphicx}
\usepackage{graphicx}% Include figure files
\usepackage{dcolumn}% Align table columns on decimal point
\usepackage{bm}% bold math
\usepackage{amsmath}
\usepackage{amsfonts}
\usepackage{amssymb}
\usepackage{color}
\usepackage{pstcol}
\usepackage[all]{xy}
\def\be{\begin{equation}}
\def\ee{\end{equation}}
\begin{document}
\title{Midisuperspace Supersymmetric Quantum Cosmology}

\author{Alfredo Mac\'{\i}as}
 \email{amac@xanum.uam.mx}
\affiliation{Departamento de F\'{\i}sica, Universidad Aut\'onoma Metropolitana--Iztapalapa,\\
A.P. 55--534, M\'exico D.F. 09340, M\'exico}
\author{Abel Camacho}
\email{acq@xanum.uam.mx}
\affiliation{Departamento de F\'{\i}sica, Universidad Aut\'onoma Metropolitana--Iztapalapa,\\
A.P. 55--534, M\'exico D.F. 09340, M\'exico}
\author{Jutta Kunz}
 \email{kunz@theorie.physik.uni-oldenburg.de}
\affiliation{Department of Physics, \\ Carl von Ossietzky
University Oldenburg, D--26111 Oldenburg, Germany}
\author{Claus L\"ammerzahl}\email{laemmerzahl@zarm.uni-bremen.de}
\affiliation{ZARM, University of Bremen, Am Fallturm, 28359
Bremen, Germany }

\date{\today}% It is always \today, today,
             %  but any date may be explicitly specified

\begin{abstract}
We investigate the canonical quantization in the framework of
$N=1$ simple supergravity for the case of a very simple
gravitational midisuperspace described by Gowdy $T^3$ cosmological
models. We consider supersymmetric quantum cosmology in the
mentioned midisuperspace, where a matrix representation for the
gravitino covector--spinor is used. The full Lorentz constraint
and its implications for the wave function of the universe are
analyzed in detail. We found that there are indeed physical states
in the midisuperspace sector of the theory in contrast to the case
of minisuperspace where there exist no physical states.

{\bf File: gowsg5.tex; 30.11.2007}
\end{abstract}


\maketitle

%**************************************************
\section{Introduction}
\label{sec1}

According to Misner \cite{mi72,ryan}, quantum cosmology is the
evolution of cosmological space--times as trajectories in the
finite dimensional sector of superspace, the so called
minisuperspace, related to the finite number of parameters that
describe $t=const.$ slices of the models and the quantum version
of such models, respectively. Taking the metric of a cosmological
model which is truncated by an enormous degree of imposed
symmetry, and plugging it into a quantization procedure cannot
give an answer that could be consider in any way as a quantum
gravity solution. What is being done in quantum cosmology, is the
assumption that one can represent a metric as a series expansion
in space--dependent modes, where the cosmological minisuperspace
model is the homogeneous mode, and the cosmological midisuperspace
model is the first non--homogeneous mode. This artificial
``freezing'' of the modes before quantization is an obvious
violation of the uncertainty principle and cannot lead to an exact
solution. However, the results of applying this untenable
quantization procedure have always seemed to predict a rather
reasonable and internally consistent behavior of the universe that
it has been difficult to believe that it does not have any
physical meaning.

After the invention of supergravity by Freedman, Nieuwenhuizen,
and Ferrara \cite{fnf76}, Teitelboim \cite{tei77a,tei77b,tei77c}
showed that this theory provides a natural classical square root
of gravity \'a la Dirac. Taking the square root of the constraints
amounts to take the square root of the corresponding quantum
equations, introducing spin in a natural way. Thereby, the total
number of constraints of the theory increases. Besides the
constraints of the original theory, there appear now new
constraints (the square roots) closing under anticommutation. The
complete set of constraints forms a graded algebra
\cite{tei77b,tei77c}. The role of the Dirac square--root will be
played by the new constraints. Furthermore, the local
supersymmetry of the action should have profound consequences upon
the resulting quantum theory, for example, the supersymmetric
constraints will provide a Dirac square root of the second--order
Wheeler--DeWitt equation governing the dynamics of the wave
function of the universe.

The classical field equations arising from the $N=1$ supergravity
Lagrangian were derived by Pilati \cite{pi78} by using the
canonical formalism. There are constraints for each of the gauge
symmetries contained in the theory, i.e., spacetime,
diffeomorphisms, local Lorentz invariance, and supersymmetry. One
important result that follows from the analysis of the field
equations is that  the Cartan relation relates the torsion tensor
and the Rarita--Schwinger gravitino field, so that it can be used
to eliminate the torsion tensor from the theory.

The canonical quantization of supergravity is performed in general
by applying Dirac's procedure for constrained systems. According
to it, quantization is performed by choosing a foliation for
spacetime, i.e., a $(3+1)$ decomposition of the canonical theory,
in which the Lagrange multipliers are the normal components
constraining the symmetry generators of the corresponding gauge
fields. Of course, all the constraints should annihilate the
ground state of the wave function. For the supergravity case,
there are three different constraints in the problem, namely, the
generators ${\cal H}_\mu$ of the translations (Hamiltonian and
diffeomorphism), the generators ${\cal J}_{\mu \nu}$ of local
Lorentz rotations and the Majorana spinor (Fermionic functions)
supersymmetric generators ${\cal S}$. The Lagrange multipliers
constraining these generators are the normal components $e^A{}_0$,
$\omega_{0 AB}$, and ${\overline \Psi}_0$, of the coframe,
connection and gravitino field, respectively.

It turns out \cite{tei77b} that the Hamiltonian constraint is
identically satisfied once the supersymmetric constraint is
fulfilled since they satisfy the relation $\left\{{\cal
S}(x),{\overline {\cal S}}(x^{\prime}) \right\} = \gamma^A\, {\cal
H}_A\,\delta(x,x^{\prime})$. Accordingly, only the Lorentz and
supersymmetric constraints are the central issue of the
quantization problem.

The gravitino field appearing in the constraints can be realized
in two different ways, namely, by differential operators or by
matrices as in the Dirac equation. In this work we will use a
matrix representation, \'a la Dirac, for the gravitino field and
since its corresponding momenta are proportional to
Rarita--Schwinger field itself, we will not rename them as it
happens in the differential operators approach \cite{mor}.

It is important to stress that general relativity, and therefore
supergravity, does not seem to possess a natural time variable,
while quantum theory relies quite heavily on a preferred time
\cite{Kuchar}. Since the nature of time in quantum gravity is not
yet clear, the classical constraints of canonical supergravity do
not contain any time parameter, after applying to them the
canonical quantization procedure. Therefore, it is needed a kind
of {\em internal time}, which is fixed by means of a gauge choice,
or by a classical solution to drive the dynamical behavior of the
resulting quantum theory \cite{maque06}.

As mentioned above, the minisuperspace is often known as the
homogeneous cosmology sector, infinitely many degrees of freedom
are artificially frozen by symmetries. This reduction is so
drastic that only a non--physical finite number of degrees of
freedom is left. The requirement of homogeneity restricts the
allowed hypersurfaces to the leaves of a privileged foliation,
which is labelled by a single {\em internal time} variable, it is
usually the volume. One can parameterize such hypersurfaces of
homogeneity by the standard Euler angles coordinates and
characterize the spatial metric uniquely by three real parameters.

The supersymmetric approach to quantum cosmology was first
introduced by Mac\'{\i}as et al. \cite{mor}, and means the study
of $N=1$ quantum supergravity models restricted to the homogeneous
minisuperspace sector of the Wheeler's superspace as direct
generalizations of standard Wheeler--De--Witt quantum cosmological
models. The standard approach to quantum cosmology consists in a
canonical quantization of homogeneous minisuperspace model, which
is obtained by imposing certain symmetry conditions on the metrics
allowed on the spacelike slices of the universe \cite{qc,ryan1}.
The dynamics of the system is governed by the Wheeler--DeWitt
equation which is a second order Klein--Gordon--like differential
constraint equation for the state function of the universe
\cite{kucryan} . The most general minisuperspace models analyzed
in the literature correspond to homogeneous and anisotropic
Bianchi type cosmological models. Since the corresponding metrics
depend only on time, the dynamics of the spacelike 3--dimensional
slices becomes trivial, unless an additional reparametrization is
performed. Usually, in the reparametrization one of the scale
factors of the Bianchi type metric, i.e., the volume  $\Omega$ of
the Misner parametrization, is fixed as {\em internal time}, as
consequence of fixing a gauge, so that the Wheeler--DeWitt
equation generates a state function of the universe which
explicitly depends on the gauge fixed {\em internal time} and on
the remaining scale factors, related to the anisotropy of such
models. It is worthwhile to stress that the volume $\Omega$ {\em
is not} a proper time parameter.

It has been found that in the framework of the minisuperspace
sector of simple supergravity approach, without having a (super)
Casimir operator, there are no physical states. Moreover, there
exists only a non--physical trivial {\em rest frame} type (bosonic
state) state function \cite{kaku}. However, the trivial ``rest
frame" type solution exist only for arbitrary Lorentz symmetry
generators \cite{ho,gc,gc2,pde,opr,maoso,mms98,mac99}.

In all the cases the failure to find physical states \cite{mamilo}
could be attributed to the fact that, due to the strong symmetry
reduction, only a finite number of degrees of freedom can be
considered, in the minisuperspace. To face this difficulty one
needs to analyze genuine field theories with an infinite number of
degrees of freedom. An option would be to consider milder symmetry
reductions which leave unaffected a specific set of true local
degrees of freedom. These are the so called midisuperspace models,
which break the homogeneity of the standard Bianchi models. The
midisuperspace models provide a canonical description of Einstein
spacetimes with a group of isometries. Symmetries remove
infinitely many degrees of freedom of the gravitational field, but
there still remain infinitely many degrees of freedom. In spite of
this simplification, the midisuperspace constraints of general
relativity are still complicated functionals of the canonical
variables, without a natural time parameter.

The simplest midisuperspace generalization of the homogeneous
minisuperspace models are the Gowdy cosmological models, since
they possess two Killing vectors, i.e., two ignorable coordinates,
reducing the problem to time (as in standard quantum cosmology)
and to one space coordinate, which completely eliminates
homogeneity and leads to a system with an infinite number of
degrees of freedom, i.e., a true field theory. Such spacetimes
have a long history in general relativity. The field equations in
this case can be shown to be equivalent to the wave equation for a
scalar field propagating in a fictitious flat (2+1)--dimensional
spacetime \cite{krameretal}. The local degrees of freedom are
contained in the scalar field. In fact, the study of
midisuperspace models and covariant field systems like string
models indicates that if there exists an {\em internal time} which
converts the old constraints of general relativity into a
Schr\"odinger equation form, such a time variable is non--local
functional of the geometric variables.

The Gowdy $T^3$ cosmological models have been analyzed in the
context of non-perturbative canonical quantization of gravity
\cite{ashpierri,mena}. The arbitrariness in the selection of a
time parameter is a problem that immediately appears in the
process of quantization. For a specific choice of time, it was
shown that there does not exist a unitary operator that could be
used to generate the corresponding quantum evolution. Therefore,
even in the case of midisuperspace models there is no natural time
parameter.

In this work we will consider the specific midisuperspace
described by Gowdy $T^3$ cosmological models \cite{gow71,gow74},
in the context of $N=1$ supergravity. The quantum constraints of
the theory are analyzed in the search of physical states.

This paper is organized as follows.  In Section \ref{sec2}, the
canonical formulation of simple supergravity $N=1$ is briefly
revisited. In \ref{sec3}, the model independent Lorentz constraint
is analyzed and explicitly solved, following closely notations and
conventions of \cite{mamilo}. In Section \ref{sec4}, the Gowdy
$T^3$ cosmological models and their main properties are reviewed.
Section \ref{sec5} is devoted to the investigation of the model
dependent supersymmetric constraint, and in \ref{sec6} we find
solutions for the state function of the universe, for both the
polarized case and for the unpolarized case. The last section
contains several final remarks.

%***************************************************************
\section{Canonical formulation of $N=1$ supergravity}
\label{sec2}

The starting point is the $(N=1)$ supergravity Lagrangian
\begin{equation}
{\cal L} = \frac{1}{2} \sqrt{-g} R - \frac{i}{2}
\varepsilon^{\lambda \mu \nu \rho} \overline{\Psi_\lambda}\gamma_5
\gamma_\mu D_{\nu} \Psi_\rho \label{lag}\, ,
\end{equation}
where
\begin{equation}
D_\nu= \partial_\nu + (1/2) \omega_{\nu AB}\sigma^{AB}
\end{equation}
is the covariant derivative and $\sigma^{AB}:=
(1/4)(\gamma^A\gamma^B - \gamma^B \gamma^A)$.

For the $\gamma^A$ matrices we use a real Majorana representation
%\begin{displaymath}
\begin{equation}
\gamma^{0}=\left( \begin{array}{cc} 0&\sigma^2\\
\sigma^2&0\end{array}\right) \, , \,\,
\gamma^{1}=\left( \begin{array}{cc} i\sigma^3&0\\
0&i\sigma^3\end{array}\right)\, , \,\,
\gamma^{2}=\left( \begin{array}{cc} 0&-\sigma^2\\
\sigma^2&0\end{array}\right)\, , \,\,
\gamma^{3}=\left( \begin{array}{cc} -i\sigma^1&0\\
0&-i\sigma^1\end{array}\right) \, , \label{realmajorana}
\end{equation} in which the
anticommutator relation $\{\gamma^A,\gamma^B\} = 2\eta^{AB}$ is
satisfied, and $\sigma^i$ are the standard Pauli matrices.
Moreover, $\gamma_5= i\gamma^0\gamma^1\gamma^2\gamma^3$. The
Rarita--Schwinger field $\Psi:=\Psi_A \omega^A$, a spinor--valued
one--form, is subject to the Majorana condition ${\overline \Psi}
= \Psi^T C$, with $C$ the charge conjugation matrix. The
vector--spinor gravitino field can be written in components form
as
\begin{equation}
\Psi_{\mu \cal{A}} = \left( \begin{array}{c}
 \psi_{\mu 1}\\
 \psi_{\mu 2}\\
 \psi_{\mu 3}\\
 \psi_{\mu 4}\end{array}\right)
\label{vecspinor}\, ,
\end{equation}
where $\mu$ is a vector index and $\cal{A}$ is a spinor index. In
this representation the Majorana condition reads ${\overline
\Psi}=-i\Psi^T \gamma^0$.

The coupling constant is set to one and the Ricci rotation
coefficients $\omega_{\nu AB}$ read
\begin{equation}
\omega_{\nu AB} = \widetilde{\omega}_{\nu AB} + K_{\nu AB}
\label{rrc} \, ,
\end{equation}
where $\widetilde{\omega}_{\nu AB}$ are the standard Levi--Civita
Ricci rotation coefficients. The contortion tensor is
 \begin{equation}
K_{\nu A B} = e_A{}^\mu e_B{}^\rho K_{\nu\mu\rho}\qquad {\rm and}
\qquad K_{\nu\mu\rho} = \frac{1}{2}\left(T_{\nu\mu\rho}-
T_{\mu\rho\nu} +T_{\rho\nu\mu} \right)\,.
\end{equation}
Greek indices from the end of the alphabet, i.e. $\lambda, \mu,
\nu, \rho, \cdots$, always range over $0,1,2,3$, Greek indices
from the beginning of the alphabet i.e. $\alpha, \beta,
\gamma,\cdots$, over $1,2,3$, and both refer to world coordinates.
Capital Latin indices, i.e. $A,\ B,...$ run over $0,1,2,3$ and
small Latin indices, i.e. $a,\ b,...$ over $1,2,3$, and are those
with respect to a local orthonormal basis.

In the case at hand the canonical variables are the covariant
spatial components of the vierbein $e^a{}_\alpha$, their conjugate
momenta $p_a{}^{\alpha}$, and the spatial covariant components of
the vector spinor $\Psi_\alpha$, defined on a generic spacelike
hypersurface. There are three different constraints in the
problem, namely, the generators ${\cal H}_\mu$ of the translations
and diffeomorphisms, the generators ${\cal J}_{\mu \nu}$ of local
Lorentz rotations and the Majorana spinor supersymmetric
generators ${\cal S}$.

The Lagrange multipliers constraining  the generators of
translations, rotations, and supersymmetry transformations are the
normal components $e^A{}_0$, $\omega_{0 AB}$, and ${\overline
\Psi}_0$, respectively, of the corresponding gauge fields
$e^A{}_\mu$, $\omega_{\mu AB}$ and ${\overline \Psi}_\mu$ with
respect to the timelike normal vector ${\bf n}$.

Therefore, the canonical form of the simple $(N=1)$ supergravity
Lagrangian (\ref{lag}) can be written as \cite{pi78}
\begin{eqnarray}
H&=& e^A{}_0\, {\cal H}_A + \frac{1}{2} \omega_{0}{}^{AB}\,{\cal
J}_{AB} + {\overline \Psi}_0\, {\cal S} \nonumber \\
 &=& N {\cal H}_\bot + N^i {\cal H}_i + \frac{1}{2} \omega_{0 AB}{\cal J}^{AB}
+ {\overline \Psi}_0 {\cal S} \label{sgham}\, ,
\end{eqnarray}
where ${\cal H}_A$, ${\cal J}_{AB}$ and ${\cal S}$ are constructed
from the canonical variables only and do not depend on the
multipliers. In the equivalent form of the canonical Lagrangian
${\cal H}_\bot$, ${\cal H}_i$ and ${\cal J}^{AB}$ are the usual
Hamiltonian, diffeomorphism, and rotational Lorentz bosonic
constraints, respectively, and ${\cal S}$ the supersymmetric
fermionic constraint. Now the lapse function $N=e_{0}{}^0$, the
shift vector $N_i=e_{i}{}^0$, \, $\omega_{0AB}$, and ${\overline
\Psi}_0$ are the corresponding Lagrange multipliers. The
supergravity generators satisfy the following algebra discovered
by Teitelboim \cite{tei77b}:
\begin{eqnarray}
\left\{{\cal S}(x),{\overline {\cal S}}(x^{\prime}) \right\} &=&
\gamma^A\,
{\cal H}_A\,\delta(x,x^{\prime}) \label{srg}\, , \\
\left[{\cal S}(x),{\cal H}_C(x^{\prime})\right] &=& \frac{1}{2}
\Sigma_{CAB}\, {\cal J}^{AB}\,\delta(x,x^{\prime}) \, ,\\
\left[{\cal S}(x),{\cal J}^{AB}(x^{\prime})\right] &=& -
\sigma^{AB}\, {\cal S}
\,\delta(x,x^{\prime}) \, ,\\
\left[{\cal H}_A (x),{\cal H}_B(x^{\prime})\right] &=& \left(-
T_{AB}{}^C {\cal H}_C +\frac{1}{2} \Omega_{ABCD}\, {\cal J}^{CD} +
{\overline H}_{AB}\, {\cal S}\right)
\,\delta(x,x^{\prime}) \label{9}\, ,\\
\left[{\cal H}_C (x),{\cal J}^{AB}(x^{\prime})\right] &=&
\left(\delta^B_C\,
{\cal H}^A - \delta^A_C\,{\cal H}^B \right)\,\delta(x,x^{\prime}) \, ,\\
\left[{\cal J}^{AB}(x),{\cal J}^{CD}(x^{\prime})\right] &=&
\left(\eta^{AC}\, {\cal J}^{BD} - \eta^{BC}\, {\cal J}^{AD} +
\eta^{BD}\, {\cal J}^{AC} - \eta^{AD}\, {\cal
J}^{BC}\right)\,\delta(x,x^{\prime}) \label{alg}\, .
\end{eqnarray}
Note that even the bosonic part is only a closed {\em soft} gauge
algebra \cite{sonius} due to the appearance of torsion and
curvature, instead of structure constants, on the right hand side.
The fields
\begin{eqnarray}
H_{AB} &=& D_A\Psi_B - D_B\Psi_A \, ,\\
\Sigma _{ABC} &=& \gamma_5 \left( \gamma_A ^* H_{BC} + \frac{1}{2}
e_{A}{}^\mu\, e_{[B \mu}\gamma_D\, {}^*H_{C]}{}^D \right)\, , \\
\Omega_{ABCD}  &=& R_{ABCD} - {\overline \Psi}_{[A}\Sigma_{B]\,CD}
\label{curvt} \, ,
\end{eqnarray}
play the role of curvature two--forms and depend on the canonical
variables of the theory. Without them, the algebra goes over into
the supersymmetry algebra of flat space \cite{wz74}.

Consequently, physical states $\vert\Psi\rangle$ in the quantum
theory have to satisfy the conditions
\begin{equation}
{\cal S} \vert\Psi\rangle=0\, , \qquad {\cal H}_A
\vert\Psi\rangle=0\, , \qquad {\cal J}_{AB} \vert\Psi\rangle = 0
\label{phst}\, .
\end{equation}
Note that the supersymmetric  constraint ${\cal S}
\vert\Psi\rangle=0$ is the ``square root" of the Hamiltonian one,
on account of (\ref{srg}), and implies ${\cal H}_A
\vert\Psi\rangle=0$, so the second condition is redundant. Thus,
we will focus only on the Lorentz ${\cal J}_{AB}$ and
supersymmetric ${\cal S}$ constraints, which are explicitly given
as follows \cite{pi78}:
\begin{eqnarray}
{\cal J}_{AB} &\equiv& p_{A}{}^{\alpha} e_{B \alpha} -
p_{B}{}^{\alpha}
e_{A \alpha} - \pi^\alpha{}_{\cal A} \sigma_{AB} \Psi_\alpha{}^{\cal A}\nonumber\\
 &=& 2p_{[A}{}^{\alpha} e_{B] \alpha} + \tau_{AB0}\nonumber\\
 &=& 2p_{[A}{}^{\alpha} e_{B]\alpha} +  \frac{1}{2} \phi_{[A{\cal A}}^T \phi_{B]}{}^{\cal A}
\label{lorentz}\, ,
\end{eqnarray}
where
\begin{equation}
\tau_{\mu\nu\lambda} = \frac{i}{4} {\overline \Psi}_{[\mu\vert}
\gamma_\lambda \Psi_{\vert \nu]} \label{spin}\, ,
\end{equation}
are the components of the spin tensor, see (8.7) of \cite{mie86},
$\phi_{A{\cal A}}$ are the desitized local gravitino components
(see Eq. (\ref{grav})), and
\begin{equation}
\pi^\alpha=\frac{i}{2} \varepsilon^{0 \alpha \delta \beta}
{\overline \Psi}_\delta \gamma_5 \gamma_\beta \label{momgra}\,
\end{equation}
is the momentum conjugate to the gravitino field. In the last step
we have  used the Majorana condition \cite{mimamo} ${\overline
\Psi}= \Psi^T C=-i \Psi^T \gamma^0$. Equivalently in terms of the
dual generators
\begin{equation}
{\cal J}_A = \frac{i}{2} \epsilon_{0ABC}{\cal J}^{BC}\qquad
\Rightarrow \qquad {\cal J}_0=0 \label{hitlerito}\, ,
\end{equation}
the Lorentz constraint reads
\begin{equation}
{\cal J}_A = \frac{i}{2} \epsilon_{0ABC}\left[2p^{[B\alpha}
e^{C]}{}_\alpha +  \frac{1}{2} \phi^{T[B}{}_{\cal A} \phi^{C]{\cal
A}}\right] \label{lorentz1}\, .
\end{equation}
It is interesting to note that, as expected due to the time
arbitrariness, the condition ${\cal J}_0=0$ implies that ${\cal
J}_{0B}\equiv 0$, therefore reducing the Lorentz constraint to
pure spatial rotations on the hypersurface of constant time.

The generator of supersymmetry reads \cite{pi78}
\begin{eqnarray}
{\cal S}=-i \epsilon^{ijk} \gamma_5 \gamma_i \nabla_j \Psi_k -
\frac{i}{2} p^\alpha{}_A \gamma^A \Psi_\alpha+\frac{1}{4}
{}^{(3)}e\gamma_\perp \psi_\alpha \bar{\psi}^\alpha
 \gamma^\beta \Psi_\beta \label{susyc}\, ,
\end{eqnarray}
where $\gamma_\perp=-N\gamma^0$, with $N$ the lapse function.

A further constraint, {\em the Cartan relation}
\begin{equation}
T_{\mu\nu\lambda}= - 4\tau_{\mu\nu\lambda}= - i {\overline
\Psi}_{[\mu\vert} \gamma_\lambda \Psi_{\vert\nu]} \label{cartan}\,
,
\end{equation}
relates the torsion tensor to the Rarita--Schwinger field and is
used to eliminate the torsion from the theory, leaving it only
with first class constraints \cite{mac96}.

It is rather convenient to use instead of the gravitino field
itself, its densitized local components
\begin{equation}
\phi_a = e\, e_a{}^\alpha \Psi_\alpha \label{grav}\, ,
\end{equation}
as the basic fields commuting with all non--spinor variables, here
$e={}^{(3)}e= \det(e_a{}^\alpha{})$. This variable was already
found to be the natural one for the gravitino field, see
\cite{dks77}. This choice suggests a matrix realization of the
$\phi_{i {\cal A}}$ obeying
\begin{equation}
\lbrace \phi_{i {\cal A}}, \phi_{j {\cal B}} \rbrace =
-\frac{i}{8}  (\gamma_j \gamma_i)_{{\cal A}{\cal B}}
\label{grav1}\, .
\end{equation}
Here ${\cal A}$ and ${\cal B}$ are spinor indices, and the
gravitational variables appear nowhere.

%****************************************************************

\section{Lorentz constraint}
\label{sec3}

We will assume the following form for the wave function of the
universe
\begin{equation}
\vert\Psi\rangle= \Psi_\mu = \left( \begin{array}{c}
 \Psi_I\\
 \Psi_{II}\\
 \Psi_{III}\\
 \Psi_{IV}\end{array}\right)
\label{wfu}\, ,
\end{equation}
Using the real Majorana representation (\ref{realmajorana}) for
the $\gamma$--matrices \cite{Niew81,kaku} as well as the
anticommuting relation (\ref{grav1}) between the components of the
gravitino field, we can write the components of the Lorentz
generator (\ref{lorentz1}) and of the supersymmetric generator
(\ref{susyc}).

It is well known that as we fix a particular basis for the
vierbein, as, for instance, the $SO(3)$ one, the Lorentz
constraint (\ref{lorentz1}) reduces to
\begin{equation}
{\cal J}_A = \frac{i}{2} \epsilon_{0ABC} \left[\frac{i}{2}
\phi_{[B{\cal A}}^T \phi_{C]}{}^{\cal A}\right]
\label{lorentz1a}\, .
\end{equation}
Therefore \begin{eqnarray}
{\cal J}_1 &=& -\frac{i}{2}\left[ \phi_{2 {\cal A}} \phi_3{}^{\cal A}
+ \phi_{3 {\cal A}} \phi_2{}^{\cal A}\right] \label{compoJ1}\, ,\\
{\cal J}_2 &=& -\frac{i}{2}\left[ \phi_{3 {\cal A}} \phi_1{}^{\cal A}
+ \phi_{1 {\cal A}} \phi_3{}^{\cal A}\right] \label{compoJ2}\, , \\
{\cal J}_3 &=& -\frac{i}{2}\left[ \phi_{1{\cal A}} \phi_2{}^{\cal
A}+ \phi_{2 {\cal A}} \phi_1{}^{\cal A}\right] \label{compoJ3}\, .
\end{eqnarray}
By means of the algebra (\ref{grav1}), which the components of the
gravitino field fulfill, we arrive at a realization of the
components of the Lorentz constraint in terms of  the standard
generators of the ordinary rotation group $O(3)$ \cite{kaku}
\begin{equation}
{\cal J}_{3}=-i\left( \begin{array}{cccc} 0&0&0&0\\
0&0&1&0\\0&-1&0&0\\0&0&0&0\end{array}\right) \, ,
{\cal J}_{2}=-i\left( \begin{array}{cccc} 0&0&0&0\\
0&0&0&-1\\0&0&0&0\\0&1&0&0\end{array}\right)\, , \,\,
{\cal J}_{1}=-i\left( \begin{array}{cccc} 0&0&0&0\\
0&0&0&0\\0&0&0&1\\0&0&-1&0\end{array}\right) \label{bizcotel}\, .
\end{equation}
Consequently, for instance the component $\Psi_{II}$ of the state
function should have four components, i.e., $\Psi_{II} =
\left(\Psi^1_{II}, \Psi^2_{II}, \Psi^3_{II}, \Psi^4_{II}\right)$,
analogously for $\Psi_{III}$, and $\Psi_{IV}$.

Let us analyze the Lorentz condition ${\cal J}_{AB}
\vert\Psi\rangle = 0$ which explicitly reads
\begin{equation}
{\cal J}_{AB}\vert\Psi\rangle = \left( \begin{array}{cccc}
 0&0&0&0\\
0&~0&~{\cal J}_{12}&~{\cal J}_{13}\\
0&~-{\cal J}_{12}&~~0&~{\cal J}_{23}\\
0&~-{\cal J}_{13}&~- {\cal J}_{23}&~~0
\end{array}\right)
\left( \begin{array}{c}
 \Psi_I\\
 \Psi_{II}\\
 \Psi_{III}\\
 \Psi_{IV}\end{array}\right)= 0
\label{lor}\, .
\end{equation}
This implies the conditions \footnote{These relations are quite
general, they do not depend on the particular Bianchi model into
consideration, cf. \cite{Niew81}}
\begin{eqnarray}
{\cal J}_{12}\Psi_{III} &=& -{\cal J}_{13}\Psi_{IV}\label{sys1}\, ,\\
{\cal J}_{12}\Psi_{II} &=& {\cal J}_{23}\Psi_{IV}\label{sys2}\, , \\
{\cal J}_{13}\Psi_{II} &=& -{\cal J}_{23}\Psi_{III} \label{sys3}\,
,
\end{eqnarray}
or equivalently, we can write the conditions
(\ref{sys1})--(\ref{sys3}) as
\begin{eqnarray}
{\cal J}_3 \Psi_{III} &=& {\cal J}_2 \Psi_{IV} \label{hsys1}\, ,\\
{\cal J}_3 \Psi_{II} &=& {\cal J}_1 \Psi_{IV} \label{hsys2}\, ,\\
{\cal J}_2 \Psi_{II} &=& {\cal J}_1 \Psi_{III} \label{hsys3}\, ,
\end{eqnarray}
respectively.

It is interesting to note that there is no condition in
(\ref{sys1})--(\ref{sys3}) or equivalently in
(\ref{hsys1})--(\ref{hsys3}) involving $\Psi_I$. By replacing the
representation (\ref{bizcotel}) into Eqs.
(\ref{hsys1})--(\ref{hsys3}), one obtains the following system of
algebraic equations for the different components of the state
function of the universe
\begin{eqnarray}
\Psi^2_{III} &=& \Psi^2_{IV}=0\,, \quad \quad
\Psi^3_{III}=-\Psi^4_{IV}\,, \label{ququrrua}\\
\Psi^3_{II} &=& \Psi^3_{IV}=0\, , \quad
\quad\Psi^2_{II}=-\Psi^4_{IV}\, , \label{ququrrub}
\\
\Psi^4_{II} &=& \Psi^4_{III}=0\,, \quad \quad
\Psi^2_{II}=-\Psi^3_{III}\label{ququrruc}\, .
\end{eqnarray}
The solution of (\ref{ququrrua})--(\ref{ququrruc}) is
straightforward and reads
\begin{equation}
\vert\Psi\rangle=\left( \begin{array}{c}\Psi_I\\
\Psi^1_{II}\\\Psi^1_{III}\\\Psi^1_{IV}\end{array}\right)
\label{statefunction}\, ,
\end{equation}
and reduces each of the $\Psi_{II}$, $\Psi_{III}$, and $\Psi_{IV}$
to only one component.

This ends the analysis of the Lorentz constraint. Notice that in
the bosonic Wigner ``rest-frame"--like solution for the state
function of the universe is a scalar with only one independent
component \cite{mamilo}.

%*****************************************************************
\section{Gowdy $T^3$ cosmological models}
\label{sec4}

Gowdy cosmological models are inhomogeneous time--dependent
solutions of Einstein's vacuum equations with compact Cauchy
spatial hypersurfaces whose topology can be either $T^3$ or
$S^1\times S^2$ \cite{gow71,gow74}. Other particular topologies
are contained in these two as special cases. Here we will focus on
$T^3$ models for which the line element can be written as
\cite{misner73}
\begin{equation}
ds^2 = e^{-\frac{\lambda}{2} + 3\tau} d\tau^2 -
e^{-\frac{\lambda}{2} - \tau} d\chi^2 - e^{2\tau} \left[e^P
(d\sigma + Q d\delta)^2 + e^{-P} d\delta^2\right]\ , \label{gle}
\end{equation} where $P$, $Q$, $\lambda$, and $\tau$ depend on the
non-ignorable coordinates $t$ and $\chi$. The spatial
hypersurfaces $(\tau = const.)$ are compact if we require that
$0\leq \chi, \sigma, \delta \leq 2\pi$. The expression in square
brackets depicts the metric on the $T^2$ subspace which is
generated by the commuting Killing vectors $\partial_\sigma$ and
$\partial_\delta$. The coordinate $\chi$ labels the different
tori.

When the Killing vectors are hypersurface orthogonal, the general
line element (\ref{gle}) becomes diagonal with $Q=0$ and the
corresponding cosmological models are called polarized. In this
last case, the subspace $T^2$ corresponds to the spatial surfaces
of a $(2+1)$ fictitious flat spacetime in which a scalar field,
represented by the metric structural function $P$, propagates
\cite{ashpierri}. The local degrees of freedom contained in the
scalar field are true gravitational degrees of freedom which
cannot be eliminated by a choice of gauge. We are thus facing a
genuine field theory which is a special case of a midisuperspace
model. Notice that the infinite number of degrees of freedom
contained in this midisuperspace model can be associated with the
inhomogeneous character of the spacetime. If we neglect the
inhomogeneities present in the model, we would obtain a
minisuperspace model with a finite number of degrees of freedom,
probably related to a Bianchi cosmological model. The general
unpolarized case $(Q\neq 0)$ also corresponds to a midisuperspace
model; however, its interpretation in terms of a dynamical scalar
field in a $(2+1)$ spacetime can not be realized.

In order to write the Gowdy line element (\ref{gle}) in ADM form
\cite{ber74} we introduce the lapse $N$ and shift functions $N_i$
as follows, c.f. \cite{misner73}
\begin{eqnarray}
N &=& g^{-1/2}\left[g^{00}\right]^{-1/2}  =\exp[\frac{1}{4} \lambda
- \frac{3}{2}\tau]\left[g^{00}\right]^{-1/2}\, , \label{lapseg}\\
N_i &=& g_{0i} = 0 \, , \label{shiftg}
\end{eqnarray}
where, as usual, $N$ and $N_i$ are gauge functions usually fixed
to $N=1$, and $N_i=0$, which implies a restriction on the time
development of the coordinates off the initial hypersurface. A
further restriction is that $\tau$ does not depend on $\chi$,
i.e.,
\begin{equation}
\frac{\partial \tau}{\partial \chi} = 0\, , \quad \Rightarrow
\quad \tau=\tau(t)\, , \quad \Rightarrow \quad \lambda=\lambda(t)
\label{gauget}\, ,
\end{equation}
is also introduced in order to reduce the configuration space of
the problem to one in which $\lambda$ has only one degree of
freedom, i.e., $\lambda = \lambda(t)$, although $P$ and $Q$ retain
their infinitely many degrees of freedom as arbitrary functions of
$\chi$ \cite{asan07}.

Therefore, Eq. (\ref{gle}) can be written as
\begin{equation}
ds^2 = N^2 d\tau^2 - e^{-\frac{\lambda}{2} - \tau} d\chi^2 -
e^{2\tau}\left[e^P\left(d\sigma + Q d\delta\right)^2 + e^{-P}
d\delta^2\right]\, .
 \label{admgle}
\end{equation}
The structure of the line element (\ref{admgle}) suggests the
following choice for the basis
\begin{equation}
\omega^0=d\tau\, , \quad \omega^1=d\chi\, , \quad
\omega^2=(d\sigma + Q d\delta)\, , \quad \omega^3= d\delta
\label{basis}\, ,
\end{equation}
in order to write the Gowdy line element in the standard ADM form,
i.e., $ds^2= N^2 d\tau^2 + g_{ij} \omega^i \omega^j$ \cite{ryan}.
Therefore, in this basis the metric (\ref{admgle}) reduces
\begin{equation}
ds^2 = N^2 d\tau^2 - e^{(-\frac{\lambda}{2} - \tau)}
\left(\omega^1\right)^2 - e^{2\tau}
\left[e^P\left(\omega^2\right)^2 + e^{-P}
\left(\omega^3\right)^2\right]\, ,
 \label{admgle1}
\end{equation}
hence, the corresponding coframe reads
\begin{equation}
e^0 = N d\tau \, , \quad e^1 =
e^{(-\frac{\lambda}{4}-\frac{\tau}{2})}\omega^1\, , \quad e^2 =
e^{(\tau+\frac{P}{2})} \omega^2 \, , \quad e^3 =
e^{(\tau-\frac{P}{2})} \omega^3 \label{coframe}\, ,
\end{equation}
and satisfies the standard orthonormality condition
$g^{\mu\nu}e^{A}{}_{\mu}e^{B}{}_{\nu}=\eta^{AB}$, with $e^A =
e^A_{\ \mu} \omega^\mu$. The dual basis to the coframe
(\ref{coframe}) reads
\begin{equation}
\Omega_0 = N^{-1} \partial_\tau\, , \quad
\Omega_1=e^{(\frac{\lambda}{4}+\frac{\tau}{2})} \omega_1\, , \quad
\Omega_2=\frac{1}{2}e^{(-\tau-\frac{P}{2})} \omega_2\, , \quad
\Omega_3=e^{(-\tau+\frac{P}{2})} \omega_3 \label{dualcoframe}\, ,
\end{equation}
where $\omega_1=\partial_\chi$, $\omega_2=\partial_\delta$, and
$\omega_3=-Q\partial_\sigma + \partial_\delta$ are the components
of the dual basis to (\ref{basis}). In the basis (\ref{coframe})
it is straightforward to calculate the connection one--form, i.e.,
$de^A=-\omega^A{}_C \wedge e^C =-\omega^A{}_{BC}\, e^B\wedge e^C$.
Hence, the only non-vanishing components of the connection
$\omega_{ABC}$ read
\begin{eqnarray}
\omega_{110}&=&-\omega_{101}=\frac{1}{2}\left[\frac{\dot{\lambda}}{2}
+\dot{\tau}\right] \, , \quad
\omega_{220}=-\omega_{202}=-\left[\frac{\dot{P}}{2} +
\dot{\tau}\right]\,
,\quad \omega_{230}=-\omega_{203}= - e^P \dot{Q} \, , \nonumber\\
\omega_{221}&=&-\omega_{212}= -
e^{\frac{\lambda}{4}+\frac{\tau}{2}} \frac{P_\chi}{2}\, , \qquad
\omega_{231}=-\omega_{213}=- e^{\frac{\lambda}{4}+\frac{\tau}{2}}
e^P Q_\chi \, , \quad
\omega_{330} = - \omega_{303}= \frac{\dot{P}}{2} - \dot{\tau} \, , \nonumber\\
\omega_{331} &=& - \omega_{313} =
e^{\frac{\lambda}{4}+\frac{\tau}{2}} \frac{P_\chi}{2}
\label{conections} \, ,
\end{eqnarray}
where the dot means time derivative. Therefore, the corresponding
covariant derivative, i.e., $\nabla_a=\Omega_a +
\frac{1}{4}\omega_{abc}\gamma^b\gamma^c$, reads
\begin{eqnarray}
\nabla_1 &=& e^{\frac{\lambda}{4}+\frac{\tau}{2}}\, \omega_1 +
\frac{1}{4}\left[\frac{\dot{\lambda}}{2} +
\dot{\tau}\right]\gamma^1
\gamma^0\, , \\
\nabla_2 &=& \frac{1}{2} e^{(-\tau-P/2)}\, \omega_2 -
\frac{1}{2}\left[\frac{\dot{P}}{2} + \dot{\tau}\right] \gamma^2
\gamma^0 - \frac{1}{2} e^P\, \dot{Q}\, \gamma^3 \gamma^0 -
e^{\frac{\lambda}{4}+\frac{\tau}{2}} \frac{P_\chi}{4} \gamma^2
\gamma^1 \nonumber \\
&-&\frac{1}{2} e^{\frac{\lambda}{4}+\frac{\tau}{2}} e^P\, Q_\chi\, \gamma^3 \gamma^1\, , \\
\nabla_3 &=& e^{(-\tau+P/2)}\, \omega_3 + \frac{1}{2}
\left[\frac{\dot{P}}{2} - \dot{\tau}\right] \gamma^3 \gamma^0 +
e^{\frac{\lambda}{4}+\frac{\tau}{2}} \frac{P_\chi}{4}\, \gamma^3
\gamma^1\, .
\end{eqnarray}
According to (\ref{grav}), the densitized local components of the
gravitino field are thus given by
\begin{equation}
\psi_1 = e^{(-2\tau)}\phi_1\, ,\quad
\psi_2=e^{(\frac{\lambda}{4}-\frac{\tau}{2} +
\frac{P}{2})}\phi_2\, , \quad
\psi_3=e^{(\frac{\lambda}{4}-\frac{\tau}{2}- \frac{P}{2})}\phi_3\,
.
\end{equation}

%*****************************************************************
\section{Supersymmetric Constraint}
\label{sec5}

Since we are considering simple $(N=1)$ supergravity, i.e., only
two supersymmetric charges, whose square vanishes, the general
expression (\ref{susyc}) for the supersymmetric constraint reduces
to
\begin{eqnarray}
{\cal S}&=& - i \epsilon^{abc} \gamma_5 \gamma_a \nabla_b\, e
e_c{}^k\Psi_k -\frac{i}{2} p^\alpha{}_A \gamma^A e^{-1}
e^a{}_\alpha \phi_a \, \nonumber \\
&=& i \left\{\left(\gamma^1 \phi_1 -\gamma^3 \phi_3 \right)
\frac{\Pi_\tau}{4} - \left(\gamma^2 \phi_2 + \gamma^3 \phi_3 -
\gamma^1 \phi_1 \right) \frac{\Pi_\lambda}{8} - \left(3\gamma^2
\phi_2 - 2\gamma^3 \phi_3\right) \frac{\Pi_P}{4}
\right. \nonumber\\
&-& \left. \left(\gamma^2 \phi_3 - \gamma^1 \gamma^2 \gamma^3
\phi_1\right) e^P \frac{\Pi_Q}{2} - \gamma^0
\gamma^1\left(\gamma^3 \phi_3 - \gamma^2 \phi_2\right)
e^{(-\frac{\lambda}{2}+\frac{\tau}{2})}
\frac{P_\chi}{4} \right.\nonumber \\
&+& \left. \gamma^0 \gamma^2\left(\gamma^3 \phi_1 - \gamma^1
\phi_3\right) e^{(-\frac{\lambda}{2}+\frac{\tau}{2})} e^P
\frac{Q_\chi}{2} \right\}\label{lapins}\, ,
\end{eqnarray}
where $\Pi_\tau$, $\Pi_\lambda$, $\Pi_P$, and $\Pi_Q$ are the
conjugated momenta, in the selected foliation, to $\tau$,
$\lambda$, $P$, and $Q$, respectively, and the subindex $\chi$
means $\frac{d}{d\chi}$.

The complete supersymmetric constraint is obtained by integrating
the $\chi$--dependence in (\ref{lapins}), i.e.,
\begin{equation}
{\mathfrak{S}}=\int_0^{2\pi}{{\cal{S}}d\chi}\, .
\end{equation}
In order to perform the integration, we expand our generalized
coordinates and their conjugated momenta in terms of the
one--dimensional complete set of functions $(\cos m\chi , \sin
m\chi)$, namely,
\begin{eqnarray}
\Pi_P &=& \Pi_{P 0}+\sum_{n=1}^{\infty}{(\Pi_{P n}
\cos{n\chi}+ \Pi_{P -n} \sin{n\chi})}\, ,\\
P &=& P_0+\sum_{n=1}^{\infty}{(P_n \cos{n\chi}+P_{-n}
\sin{n\chi})}\, ,\\
\Pi_Q &=& \Pi_{Q 0}+\sum_{n=1}^{\infty}{(\Pi_{Q n}
\cos{n\chi}+ \Pi_{Q -n} \sin{n\chi})}\, ,\\
Q &=& Q_0+\sum_{n=1}^{\infty}{(Q_n \cos{n\chi}+ Q_{-n}
\sin{n\chi})}\,,
\end{eqnarray}
this implies that
\begin{eqnarray}
\int_0^{2\pi} \Pi_P\, d\chi &=& \int_0^{2\pi} \left[ \Pi_{P
0}+\sum_{n=1}^{\infty}{(\Pi_{P n} \cos{n\chi}+ \Pi_{P -n}
\sin{n\chi})} \right]\, d\chi = 2\pi \Pi_{P0}\, , \\
\int_0^{2\pi} P_\chi\,d\chi &=& P \Bigg\vert_0^{2\pi} = \left[
P_0+\sum_{n=1}^{\infty}{(P_n \cos{n\chi}+P_{-n}
\sin{n\chi})}\right] \Bigg\vert_0^{2\pi}=P_0 \label{pcero}\, .
\end{eqnarray}
Therefore, assuming the condition (\ref{pcero}) for $P$,
\begin{eqnarray}
\int_0^{2\pi}  e^P \Pi_Q\, d\chi &=&  e^{P_0} \int_0^{2\pi} \left[
\Pi_{Q 0}+\sum_{n=1}^{\infty}{(\Pi_{Q n} \cos{n\chi}+ \Pi_{Q -n}
\sin{n\chi})} \right]\, d\chi = 2\pi  e^{P_0} \Pi_{Q 0}\, , \\
\int_0^{2\pi}  e^P Q_\chi\,d\chi &=&  e^{P_0} Q
\Bigg\vert_0^{2\pi} = e^{P_0} \left[ Q_0+\sum_{n=1}^{\infty}{(Q_n
\cos{n\chi} + Q_{-n} \sin{n\chi})}\right] \Bigg\vert_0^{2\pi}=
 e^{P_0} Q_0\, .
\end{eqnarray}

Hence, the final form of the supersymmetric constraint
(\ref{lapins}) reads
\begin{eqnarray}
{\mathfrak{S}} &=& \frac{i}{4} \left\{\left(\gamma^1 \phi_1
-\gamma^3 \phi_3 \right) \Pi_\tau - \left(\gamma^2 \phi_2 +
\gamma^3 \phi_3 - \gamma^1 \phi_1 \right)\frac{\Pi_\lambda}{2} -
2\left(3\gamma^2 \phi_2 -
2\gamma^3 \phi_3\right) \Pi_{P_0} \right. \nonumber\\
&-& \left. 4\left(\gamma^2 \phi_3 - \gamma^1 \gamma^2 \gamma^3
\phi_1\right) e^{P_0} \Pi_{Q_0} - \gamma^0 \gamma^1\left(\gamma^3
\phi_3 - \gamma^2 \phi_2\right) e^{(-\frac{\lambda}{2}+\frac{\tau}{2})} P_0 \right. \nonumber \\
&-& \left. 2\gamma^0 \gamma^2\left(\gamma^3 \phi_1 - \gamma^1
\phi_3\right) e^{(-\frac{\lambda}{2}+\frac{\tau}{2})} e^{P_0} Q_0
\right\}\label{csusyc}\, ,
\end{eqnarray}
where a constant factor has been included in a redefinition of all
integrated quantities.

%******************************************************************
\section{Physical states}
\label{sec6}

In order to quantize the problem that we have outlined above, we
will convert $\Pi_\tau$, $\Pi_\lambda$, $\Pi_{P_0}$, $\Pi_{Q_0}$,
$P_0$, and $Q_0$ into operators $i\frac{\delta}{\delta \tau}$,
$i\frac{\delta}{\delta \lambda}$, $i\frac{\delta}{\delta P_0}$,
$\frac{\delta}{\delta Q_0}$, ${\widehat{P}}_0$, ${\widehat{Q}}_0$,
respectively. They act on the state function of the universe
$\Psi$ and the supersymmetric constraint $\mathfrak{S}$, Eq.
(\ref{csusyc}), becomes also an operator which, according to the
Dirac canonical quantization procedure, should annihilate the
state function of the universe, i.e.,
\begin{equation}
\widehat{\mathfrak{S}}\vert\Psi\rangle=0 \label{susycon}\, .
\end{equation}

The solutions to the supersymmetric constraint, Eq.
(\ref{susycon}), for the state function $\Psi$ given by
(\ref{statefunction}), as result of solving the Lorentz
constraint, are known as physical states of the theory.

Since $\{\widehat{\mathfrak{S}}_{\cal
A},\widehat{\mathfrak{S}}_{\cal B}\}= 0$, for ${\cal A} \neq {\cal
B}$ we can take each $\widehat{\mathfrak{S}}_{\cal A}$ to operate
in orthogonal subspaces, and we can write
$\widehat{\mathfrak{S}}\vert\Psi\rangle=0$ in the form
\begin{equation}
\left( \begin{array}{cccc}
 \widehat{\mathfrak{S}}_1&0&0&0\\
0&~\widehat{\mathfrak{S}}_2&~0&~0\\
0&~0&~\widehat{\mathfrak{S}}_3&~0\\
0&~0&~0&~\widehat{\mathfrak{S}}_4
\end{array}\right)
\left( \begin{array}{c}
 \Psi_I\\
 \Psi_{II}\\
 \Psi_{III}\\
 \Psi_{IV}\end{array}\right)= 0
\label{susyc2}\, ,
\end{equation}
where each of the $\widehat{\mathfrak{S}}_{\cal A}$ will be a
matrix operator of the smallest rank possible that produces the
appropriate algebra for $\widehat{\mathfrak{S}}$.
%*********************************************************
\subsection{The polarized case $Q=0$}
\label{subsec1}

The polarized case is obtained from (\ref{csusyc}) by setting the
metric structure function $Q=0$, i.e.,
\begin{eqnarray}
\widehat{\mathfrak{S}} &=& \frac{i}{4} \left\{i\left(\gamma^1
\phi_1 -\gamma^3 \phi_3 \right) \frac{\delta}{\delta \tau} +
\frac{i}{2} \left(\gamma^2 \phi_2 + \gamma^3 \phi_3 - \gamma^1
\phi_1 \right) \frac{\delta}{\delta \lambda} + 2i \left(3\gamma^2
\phi_2 - 2\gamma^3 \phi_3\right)\frac{\delta}{\delta P_0} \right. \nonumber \\
&-&  \left. \gamma^0 \gamma^1\left(\gamma^3 \phi_3 - \gamma^2
\phi_2\right) e^{(-\frac{\lambda}{2}+\frac{\tau}{2})}
{\widehat{P}}_0 \right\}\, .
\end{eqnarray}
The operator $\widehat{\mathfrak{S}}$ has four spinor components:
\begin{eqnarray}
\widehat{\mathfrak{S}}_1 &=& i\left(-\phi_{11} - \phi_{32}\right)
\frac{\delta}{\delta \tau} + \frac{i}{2} \left(-\phi_{24} +
\phi_{32} + \phi_{11}\right) \frac{\delta}{\delta \lambda} + 2i
\left(- 3\phi_{24} - 2\phi_{32}\right) \frac{\delta}{\delta P_0} \nonumber \\
&+& \left(\phi_{21} + \phi_{33}\right)
e^{(-\frac{\lambda}{2}+\frac{\tau}{2})} {\widehat{P}}_0 \label{s1}\, , \\
\widehat{\mathfrak{S}}_2 &=& i\left(\phi_{12} - \phi_{31}\right)
\frac{\delta}{\delta \tau} + \frac{i}{2} \left(\phi_{23} +
\phi_{31} - \phi_{12}\right) \frac{\delta}{\delta \lambda} + 2i
\left(3\phi_{23} - 2\phi_{31}\right) \frac{\delta}{\delta P_0} \nonumber \\
&+& \left(-\phi_{22} + \phi_{34}\right)
e^{(-\frac{\lambda}{2}+\frac{\tau}{2})} {\widehat{P}}_0 \label{s2}\, , \\
\widehat{\mathfrak{S}}_3 &=& i\left(-\phi_{13} - \phi_{34}\right)
\frac{\delta}{\delta \tau} + \frac{i}{2} \left(\phi_{22} +
\phi_{34} + \phi_{13}\right) \frac{\delta}{\delta \lambda} + 2i
\left(3\phi_{22} - 2\phi_{34}\right) \frac{\delta}{\delta P_0} \nonumber \\
&+& \left(- \phi_{23} - \phi_{31}\right)
e^{(-\frac{\lambda}{2}+\frac{\tau}{2})} {\widehat{P}}_0 \label{s3}\, , \\
\widehat{\mathfrak{S}}_4 &=& i\left(\phi_{14} - \phi_{33}\right)
\frac{\delta}{\delta \tau} + \frac{i}{2} \left(- \phi_{21} +
\phi_{33} - \phi_{14}\right) \frac{\delta}{\delta \lambda} + 2i
\left(- 3\phi_{21} - 2\phi_{33}\right) \frac{\delta}{\delta P_0} \nonumber \\
&+& \left(\phi_{24} - \phi_{32}\right)
e^{(-\frac{\lambda}{2}+\frac{\tau}{2})} {\widehat{P}}_0
\label{s4}\, ,
\end{eqnarray}
the components (\ref{s1})--(\ref{s4}) of the supersymmetric
constraint $\widehat{\mathfrak{S}}$ can be written in compact form
as
\begin{eqnarray}
\widehat{\mathfrak{S}}_{\cal A} &=&  i M_{{\cal A}1}
\frac{\delta}{\delta \tau} + i M_{{\cal A}2} \frac{\delta}{\delta
\lambda} + i M_{{\cal A}3} \frac{\delta}{\delta P_0} + M_{{\cal
A}4} e^{(-\frac{\lambda}{2}+\frac{\tau}{2})} {\widehat{P}}_0 \, ,
\end{eqnarray}
or equivalently
\begin{eqnarray}
\widehat{\mathfrak{S}}_{\cal A} &=&  i \Gamma^1
\frac{\delta}{\delta \tau} + i \Gamma^2 \frac{\delta}{\delta
\lambda} + i \Gamma^3 \frac{\delta}{\delta P_0} + \Gamma^4
e^{(-\frac{\lambda}{2}+\frac{\tau}{2})} {\widehat{P}}_0 \, .
\end{eqnarray}
As can be easily seen, we need to find a matrix realization
consisting of a set of four independent matrices satisfying the
algebra $\{\Gamma^A,\Gamma^B\}=0$, for $A \neq B=1,\cdots,4$. In
order to solve the equations $\widehat{\mathfrak{S}}_{\cal A}
\Psi_{\cal A}=0$ for the polarized case, we use the following $4
\times 4$ matrix realization of the $\Gamma^A$ matrices:
\begin{equation}
\Gamma^1=\left( \begin{array}{cccc}
0&0&0&i\\
0&~0&~i&~0\\
0&~i&~0&~0\\
i&~0&~0&~0
\end{array}\right)\, , \quad
\Gamma^2=\left( \begin{array}{cccc}
0&0&0&-i\\
0&~0&~i&~0\\
0&~-i&~0&~0\\
i&~0&~0&~0
\end{array}\right)\, ,
\end{equation}
\begin{equation}
\Gamma^3=\left( \begin{array}{cccc}
0&0&1&0\\
0&~0&~0&~-1\\
1&~0&~0&~0\\
0&~-1&~0&~0
\end{array}\right)\, , \quad
\Gamma^4=\left( \begin{array}{cccc}
0&0&-i&0\\
0&~0&~0&~-i\\
i&~0&~0&~0\\
0&~i&~0&~0
\end{array}\right)\, ,
\end{equation}
This choice implies that each $\Psi_{\cal A}$ splits itself into a
four components object. Therefore, the supersymmetric condition
reduces to the following set of equations
\begin{eqnarray}
i\left[\frac{\delta}{\delta {\widehat{P}}_0} -
e^{(-\frac{\lambda}{2}+\frac{\tau}{2})} {\widehat{P}}_0 \right]
\Psi_{{\cal A}3} - \left[\frac{\delta}{\delta \tau} -
\frac{\delta}{\delta \lambda}\right] \Psi_{{\cal A}4} &=&0 \, , \\
-\left[\frac{\delta}{\delta \tau} + \frac{\delta}{\delta
\lambda}\right] \Psi_{{\cal A}3} - i\left[\frac{\delta}{\delta
{\widehat{P}}_0} + e^{(-\frac{\lambda}{2}+\frac{\tau}{2})}
{\widehat{P}}_0 \right] \Psi_{{\cal A}4} &=& 0 \, , \\
i\left[\frac{\delta}{\delta {\widehat{P}}_0} +
e^{(-\frac{\lambda}{2}+\frac{\tau}{2})} {\widehat{P}}_0 \right]
\Psi_{{\cal A}1} - \left[\frac{\delta}{\delta \tau} -
\frac{\delta}{\delta \lambda}\right] \Psi_{{\cal A}2} &=&0 \, , \\
-\left[\frac{\delta}{\delta \tau} + \frac{\delta}{\delta
\lambda}\right] \Psi_{{\cal A}1} - i\left[\frac{\delta}{\delta
{\widehat{P}}_0} - e^{(-\frac{\lambda}{2}+\frac{\tau}{2})}
{\widehat{P}}_0 \right] \Psi_{{\cal A}2} &=& 0 \, ,
\end{eqnarray}
or equivalently
\begin{eqnarray}
\left[\frac{\delta}{\delta \tau} + \frac{\delta}{\delta
\lambda}\right] \Psi_{{\cal A}1} &=& 0\, , \quad
\left[\frac{\delta}{\delta {\widehat{P}}_0} +
e^{(-\frac{\lambda}{2}+\frac{\tau}{2})} {\widehat{P}}_0 \right]
\Psi_{{\cal A}1} = 0\, ,\label{e1}\\
\left[\frac{\delta}{\delta \tau} - \frac{\delta}{\delta
\lambda}\right] \Psi_{{\cal A}2} &=&0\, , \quad
\left[\frac{\delta}{\delta {\widehat{P}}_0} -
e^{(-\frac{\lambda}{2}+\frac{\tau}{2})} {\widehat{P}}_0 \right]
\Psi_{{\cal A}2} = 0\, , \label{e2}\\
\left[\frac{\delta}{\delta \tau} + \frac{\delta}{\delta
\lambda}\right] \Psi_{{\cal A}3}&=& 0 \, , \quad
\left[\frac{\delta}{\delta {\widehat{P}}_0} -
e^{(-\frac{\lambda}{2}+\frac{\tau}{2})} {\widehat{P}}_0 \right]
\Psi_{{\cal A}3} = 0 \, , \label{e3}\\
\left[\frac{\delta}{\delta \tau} - \frac{\delta}{\delta
\lambda}\right] \Psi_{{\cal A}4} &=&0 \, , \quad
\left[\frac{\delta}{\delta {\widehat{P}}_0} +
e^{(-\frac{\lambda}{2}+\frac{\tau}{2})} {\widehat{P}}_0 \right]
\Psi_{{\cal A}4} = 0 \label{e4}\, .
\end{eqnarray}
It is straightforward to see that only Eqs. (\ref{e1}) and
(\ref{e3}) can be consistently solved. Therefore, the physical
state reads
\begin{eqnarray}
\Psi_{{\cal A}1} &=& \Psi_{{\cal A}10}\exp[m(\lambda - \tau)]
\exp[-e^{(-\frac{\lambda}{2}+\frac{\tau}{2})}
\frac{{\widehat{P}}^2_0}{2}] \, , \\
\Psi_{{\cal A}2} &=& 0 \, , \\
\Psi_{{\cal A}3} &=& \Psi_{{\cal A}30}\exp[m(\lambda - \tau)]
\exp[e^{(-\frac{\lambda}{2}+\frac{\tau}{2})}
\frac{{\widehat{P}}^2_0}{2}] \, , \\
\Psi_{{\cal A}4} &=& 0\, ,
\end{eqnarray}
or equivalently
\begin{equation}
\Psi_{{\cal A}}=\exp[m(\lambda - \tau)]\left( \begin{array}{c}
\Psi_{{\cal A}10}\exp[-e^{(-\frac{\lambda}{2}+\frac{\tau}{2})}
\frac{{\widehat{P}}^2_0}{2}]\\
0 \\
\Psi_{{\cal A}30}\exp[e^{(-\frac{\lambda}{2}+\frac{\tau}{2})}
\frac{{\widehat{P}}^2_0}{2}]\\
0\\
\end{array}\right) \label{sol2}\, ,
\end{equation}
where $\Psi_{{\cal A}10}$ and $\Psi_{{\cal A}30}$ are integration
constants and $m$ is a separation constant.
%********************************************************
\subsection{The unpolarized case $Q \neq 0$}
\label{subsec2}

Let us now consider the general case of the supersymmetric
constraint for the Gowdy $T^3$ cosmological models
\begin{eqnarray}
\widehat{\mathfrak{S}} &=& \frac{i}{4} \left\{\left(\gamma^1
\phi_1 -\gamma^3 \phi_3 \right) \frac{\delta}{\delta \tau} -
\frac{i}{2} \left(\gamma^2 \phi_2 + \gamma^3 \phi_3 - \gamma^1
\phi_1 \right) \frac{\delta}{\delta \lambda} - 2i \left(3\gamma^2
\phi_2 - 2\gamma^3 \phi_3\right) \frac{\delta}{\delta P_0} \right. \nonumber \\
&-& \left. \gamma^0 \gamma^1\left(\gamma^3 \phi_3 - \gamma^2
\phi_2\right) e^{(-\frac{\lambda}{2}+\frac{\tau}{2})}
{\widehat{P}}_0 - 4i \left(\gamma^2 \phi_3 - \gamma^1 \gamma^2
\gamma^3\phi_1\right) e^{{\widehat{P}}_0} \frac{\delta}{\delta Q_0} \right. \nonumber \\
&-& \left. 2 \gamma^0 \gamma^2\left(\gamma^3 \phi_1 - \gamma^1
\phi_3\right) e^{(-\frac{\lambda}{2}+\frac{\tau}{2})}
e^{{\widehat{P}}_0} {\widehat{Q}}_0 \right\}\, .
\end{eqnarray}

\begin{figure}[h!]
\setlength{\unitlength}{1cm}
\begin{picture}(15,12)
%\put(-1,0){\epsfig{file=Fig2l0.5new.eps,width=8cm}}
\put(0,0){
\includegraphics[width=70mm,angle=0,keepaspectratio]{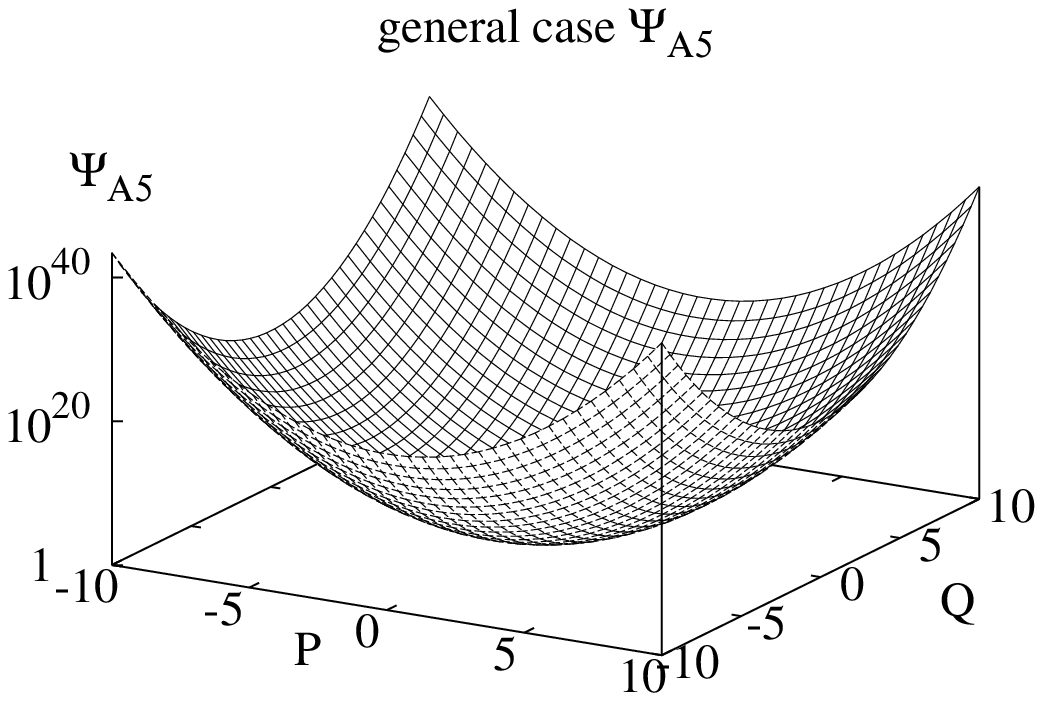}
} \put(7,0){
\includegraphics[width=70mm,angle=0,keepaspectratio]{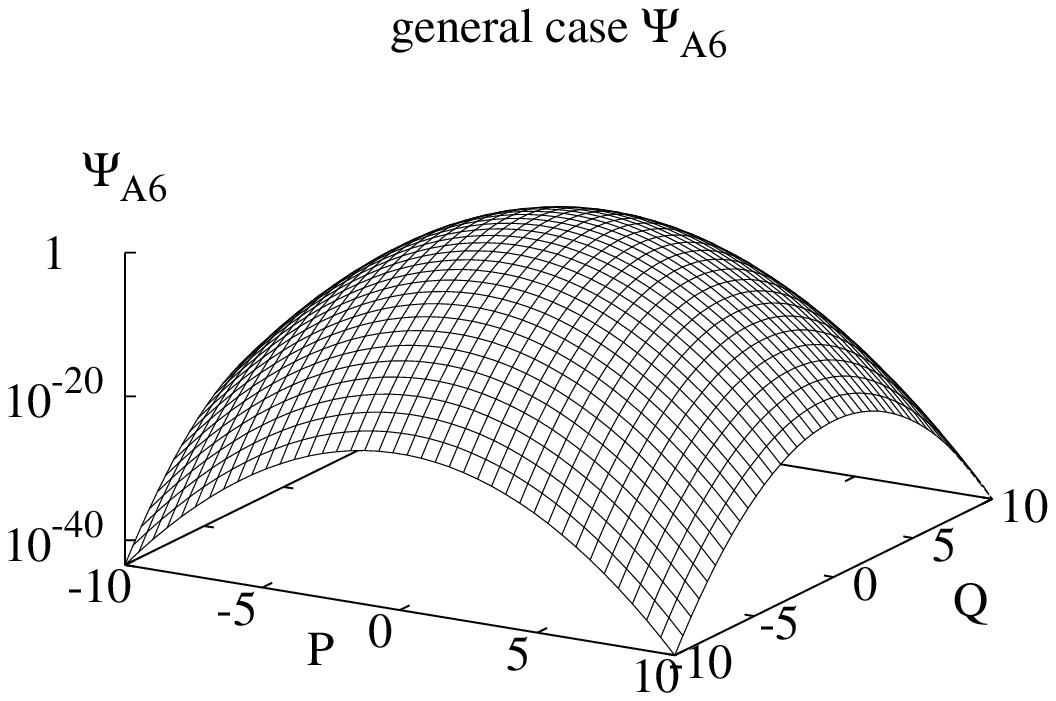}
} \put(0,6){
\includegraphics[width=70mm,angle=0,keepaspectratio]{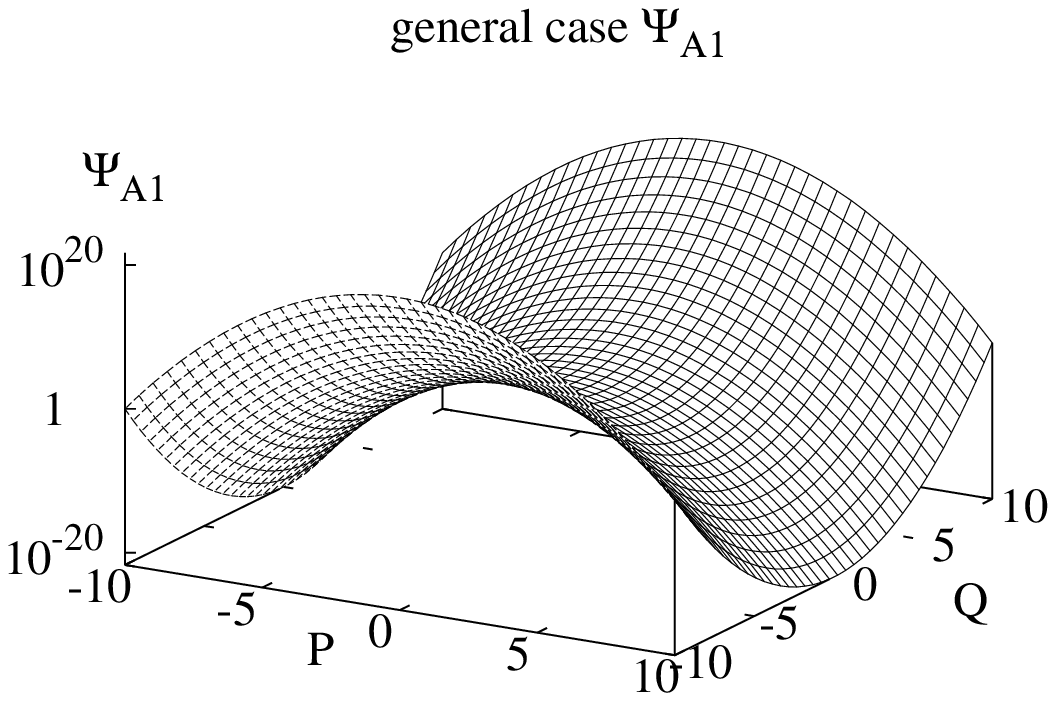}
} \put(7,6){
\includegraphics[width=70mm,angle=0,keepaspectratio]{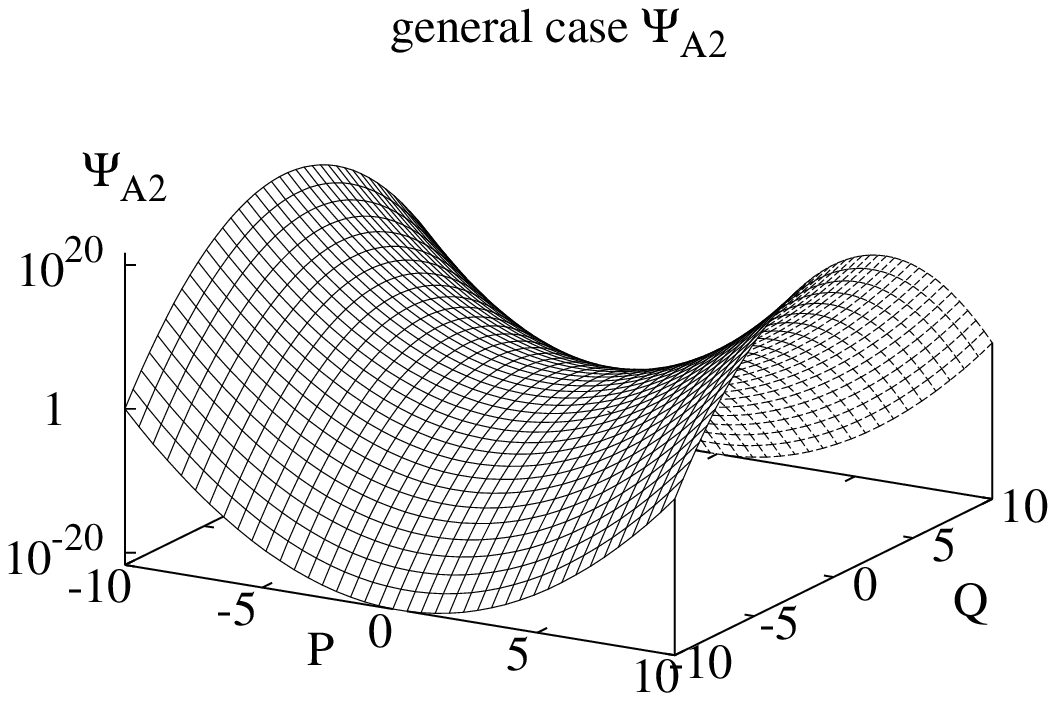}
}
\end{picture}
\caption{ Behavior of the non--vanishing components of the state
function of the universe $\Psi_{{\cal A}i}$ with respect to $P_0$
and $Q_0$ for the special solution (\ref{sol2g}). We fixed $\tau$
and $\lambda$ to a constant value.} \label{solgenplot}
\end{figure}

As before, the operator $\widehat{\mathfrak{S}}$ has four spinor
components:
\begin{eqnarray}
\widehat{\mathfrak{S}}_1 &=& i\left(-\phi_{11} - \phi_{32}\right)
\frac{\delta}{\delta \tau} + \frac{i}{2}\left(- \phi_{24} +
\phi_{32} + \phi_{11}\right) \frac{\delta}{\delta \lambda} +
2i\left(- 3\phi_{24} - 2\phi_{32}\right) \frac{\delta}{\delta P_0} \nonumber \\
&+& \left(\phi_{21} + \phi_{33}\right)
e^{(-\frac{\lambda}{2}+\frac{\tau}{2})} {\widehat{P}}_0 +
4i\left(-\phi_{13} + \phi_{34}\right) e^{{\widehat{P}}_0}
\frac{\delta}{\delta Q_0} \nonumber\\
&+& 2\left(-\phi_{31} - \phi_{12}\right)
e^{(-\frac{\lambda}{2}+\frac{\tau}{2})} e^{{\widehat{P}}_0}
{\widehat{Q}}_0 \label{s1g}\, , \\
\widehat{\mathfrak{S}}_2 &=& i\left(\phi_{12} - \phi_{31}\right)
\frac{\delta}{\delta \tau} + \frac{i}{2}\left(\phi_{23} +
\phi_{31} - \phi_{12}\right) \frac{\delta}{\delta \lambda} + 2i
\left(3\phi_{23} - 2\phi_{31}\right) \frac{\delta}{\delta P_0} \nonumber \\
&+& \left(- \phi_{22} + \phi_{34}\right)
e^{(-\frac{\lambda}{2}+\frac{\tau}{2})}{\widehat{P}}_0 +
4i\left(-\phi_{14} - \phi_{33}\right) e^{{\widehat{P}}_0}
\frac{\delta}{\delta Q_0} \nonumber \\
&+& 2\left(\phi_{32} - \phi_{11}\right)
e^{(-\frac{\lambda}{2}+\frac{\tau}{2})} e^{{\widehat{P}}_0}
{\widehat{Q}}_0 \label{s2g}\, , \\
\widehat{\mathfrak{S}}_3 &=& i\left(-\phi_{13} - \phi_{34}\right)
\frac{\delta}{\delta \tau} + \frac{i}{2}\left(\phi_{22} +
\phi_{34} + \phi_{13}\right) \frac{\delta}{\delta \lambda} + 2i
\left(3\phi_{22} - 2\phi_{34}\right) \frac{\delta}{\delta P_0} \nonumber \\
&+& \left(- \phi_{23} - \phi_{31}\right)
e^{(-\frac{\lambda}{2}+\frac{\tau}{2})} {\widehat{P}}_0 +
4i\left(\phi_{11} - \phi_{32}\right) e^{{\widehat{P}}_0}
\frac{\delta}{\delta Q_0} \nonumber \\
&+& 2\left(\phi_{33} + \phi_{14}\right)
e^{(-\frac{\lambda}{2}+\frac{\tau}{2})} e^{{\widehat{P}}_0}
{\widehat{Q}}_0 \label{s3g}\, , \\
\widehat{\mathfrak{S}}_4 &=& i\left(\phi_{14} - \phi_{33}\right)
\frac{\delta}{\delta \tau} + \frac{i}{2}\left(-\phi_{21} +
\phi_{33} - \phi_{14}\right) \frac{\delta}{\delta \lambda} +
2i\left(-3\phi_{21} - 2\phi_{33}\right) \frac{\delta}{\delta P_0} \nonumber \\
&+& \left(\phi_{24} - \phi_{32}\right)
e^{(-\frac{\lambda}{2}+\frac{\tau}{2})} {\widehat{P}}_0 +
4i\left(\phi_{12} + \phi_{31}\right) e^{{\widehat{P}}_0}
\frac{\delta}{\delta Q_0} \nonumber \\
&+& 2 \left(-\phi_{34} + \phi_{13}\right)
e^{(-\frac{\lambda}{2}+\frac{\tau}{2})} e^{{\widehat{P}}_0}
{\widehat{Q}}_0 \label{s4g}\, ,
\end{eqnarray}
once again, the components (\ref{s1g})--(\ref{s4g}) of the
supersymmetric constraint $\widehat{\mathfrak{S}}$ can be written
in compact form as
\begin{eqnarray}
\widehat{\mathfrak{S}}_{\cal A} &=&  M_{{\cal A}1}
\frac{\delta}{\delta \tau} + iM_{{\cal A}2} \frac{\delta}{\delta
\lambda} + i M_{{\cal A}3} \frac{\delta}{\delta P_0} + M_{{\cal
A}4} e^{(-\frac{\lambda}{2}+\frac{\tau}{2})} {\widehat{P}}_0 + i
M_{{\cal A}5} e^{{\widehat{P}}_0} \frac{\delta}{\delta Q_0}
\nonumber\\
&+& M_{{\cal A}6} e^{(-\frac{\lambda}{2}+\frac{\tau}{2})}
e^{{\widehat{P}}_0} {\widehat{Q}}_0 \, ,
\end{eqnarray}
or equivalently
\begin{eqnarray}
\widehat{\mathfrak{S}}_{\cal A} &=&  i\Gamma^1
\frac{\delta}{\delta \tau} + i\Gamma^2 \frac{\delta}{\delta
\lambda} + i\Gamma^3 \frac{\delta}{\delta P_0} + \Gamma^4
e^{(-\frac{\lambda}{2}+\frac{\tau}{2})} {\widehat{P}}_0 + i
\Gamma^5 e^{{\widehat{P}}_0}\frac{\delta}{\delta Q_0}  \nonumber \\
&+& \Gamma^6 e^{(-\frac{\lambda}{2}+\frac{\tau}{2})}
e^{{\widehat{P}}_0} {\widehat{Q}}_0 \, .
\end{eqnarray}
It is straightforward to see that we need to find a matrix
realization consisting of a set of six independent matrices
satisfying the algebra $\{\Gamma^A,\Gamma^B\}=0$, for $A \neq
B=1,\cdots,6$. In order to solve the equations
$\widehat{\mathfrak{S}}_{\cal A} \Psi_{\cal A}=0$ for the
unpolarized case. We use the following $8 \times 8$ matrix
realization of the $\Gamma^A$ matrices:
\begin{equation}
\Gamma^1=\left( \begin{array}{cccccccc}
0&0&0&0&0&0&0&-1\\
0&0&0&0&0&0&1&~0\\
0&0&0&0&0&-1&0&0\\
0&0&0&0&1&0&0&0\\
0&0&0&-1&0&0&0&0\\
0&0&1&0&0&0&0&0\\
0&-1&0&0&0&0&0&0\\
1&0&0&0&0&0&0&0
\end{array}\right)\, , \quad
\Gamma^2=\left( \begin{array}{cccccccc}
0&0&0&0&0&0&0&1\\
0&0&0&0&0&0&-1&0\\
0&0&0&0&0&-1&0&0\\
0&0&0&0&1&0&0&0\\
0&0&0&1&0&0&0&0\\
0&0&-1&0&0&0&0&0\\
0&-1&0&0&0&0&0&0 \\
1&0&0&0&0&0&0&0
\end{array}\right)\, ,
\end{equation}
\begin{equation}
\Gamma^3=\left( \begin{array}{cccccccc}
0&0&0&0&1&0&0&0\\
0&0&0&0&0&-1&0&0\\
0&0&0&0&0&0&1&0\\
0&0&0&0&0&0&0&-1\\
1&0&0&0&0&0&0&0\\
0&-1&0&0&0&0&0&0\\
0&0&1&0&0&0&0&0\\
0&0&0&-1&0&0&0&0
\end{array}\right)\, , \quad
\Gamma^4=\left( \begin{array}{cccccccc}
0&0&0&0&-i&0&0&0\\
0&0&0&0&0&-i&0&0\\
0&0&0&0&0&0&-i&0\\
0&0&0&0&0&0&0&-i\\
i&0&0&0&0&0&0&0\\
0&i&0&0&0&0&0&0\\
0&0&i&0&0&0&0&0\\
0&0&0&i&0&0&0&0
\end{array}\right)\, ,
\end{equation}
\begin{equation}
\Gamma^5=\left( \begin{array}{cccccccc}
0&0&0&0&0&-1&0&0\\
0&0&0&0&-1&0&0&0\\
0&0&0&0&0&0&0&-1\\
0&0&0&0&0&0&-1&0\\
0&-1&0&0&0&0&0&0\\
-1&0&0&0&0&0&0&0\\
0&0&0&-1&0&0&0&0\\
0&0&-1&0&0&0&0&0
\end{array}\right)\, , \quad
\Gamma^6=\left( \begin{array}{cccccccc}
0&0&0&0&0&-i&0&0\\
0&0&0&0&i&0&0&0\\
0&0&0&0&0&0&0&i\\
0&0&0&0&0&0&-i&0\\
0&-i&0&0&0&0&0&0\\
i&0&0&0&0&0&0&0\\
0&0&0&i&0&0&0&0\\
0&0&-i&0&0&0&0&0
\end{array}\right)\, ,
\end{equation}
This choice implies that each $\Psi_{\cal A}$ splits itself into
an eight components object. Therefore, the supersymmetric
condition $\widehat{\mathfrak{S}}_{\cal A} \Psi_{\cal A}=0$
reduces to the following set of equations
\begin{eqnarray}
\left[\frac{\delta}{\delta \tau} + \frac{\delta}{\delta
\lambda}\right] \Psi_{{\cal A}1} &=& 0\, , \,
\left[\frac{\delta}{\delta {\widehat{P}}_0} +
e^{(-\frac{\lambda}{2}+\frac{\tau}{2})} {\widehat{P}}_0 \right]
\Psi_{{\cal A}1} = 0\, , \, \left[\frac{\delta}{\delta Q_0} -
e^{(-\frac{\lambda}{2}+\frac{\tau}{2})}
{\widehat{Q}}_0\right]\Psi_{{\cal A}1} = 0\label{e1g}\\
\left[\frac{\delta}{\delta \tau} + \frac{\delta}{\delta
\lambda}\right] \Psi_{{\cal A}2} &=&0\, , \,
\left[\frac{\delta}{\delta {\widehat{P}}_0} -
e^{(-\frac{\lambda}{2}+\frac{\tau}{2})} {\widehat{P}}_0 \right]
\Psi_{{\cal A}2} = 0\, , \, \left[\frac{\delta}{\delta Q_0} +
e^{(-\frac{\lambda}{2}+\frac{\tau}{2})}
{\widehat{Q}}_0\right]\Psi_{{\cal A}2}= 0\label{e2g}\\
\left[\frac{\delta}{\delta \tau} - \frac{\delta}{\delta
\lambda}\right] \Psi_{{\cal A}3}&=& 0 \, , \,
\left[\frac{\delta}{\delta {\widehat{P}}_0} +
e^{(-\frac{\lambda}{2}+\frac{\tau}{2})} {\widehat{P}}_0 \right]
\Psi_{{\cal A}3} = 0 \, , \, \left[\frac{\delta}{\delta Q_0} +
e^{(-\frac{\lambda}{2}+\frac{\tau}{2})}
{\widehat{Q}}_0\right]\Psi_{{\cal A}3} = 0\label{e3g}\\
\left[\frac{\delta}{\delta \tau} - \frac{\delta}{\delta
\lambda}\right] \Psi_{{\cal A}4} &=&0 \, , \,
\left[\frac{\delta}{\delta {\widehat{P}}_0} -
e^{(-\frac{\lambda}{2}+\frac{\tau}{2})} {\widehat{P}}_0 \right]
\Psi_{{\cal A}4} = 0 \, , \, \left[\frac{\delta}{\delta Q_0} -
e^{(-\frac{\lambda}{2}+\frac{\tau}{2})}
{\widehat{Q}}_0\right]\Psi_{{\cal A}4}= 0\label{e4g}\,  \\
\left[\frac{\delta}{\delta \tau} + \frac{\delta}{\delta
\lambda}\right] \Psi_{{\cal A}5} &=& 0\, , \,
\left[\frac{\delta}{\delta {\widehat{P}}_0} -
e^{(-\frac{\lambda}{2}+\frac{\tau}{2})} {\widehat{P}}_0 \right]
\Psi_{{\cal A}5} = 0\, , \, \left[\frac{\delta}{\delta Q_0} -
e^{(-\frac{\lambda}{2}+\frac{\tau}{2})}
{\widehat{Q}}_0\right]\Psi_{{\cal A}5} = 0\label{e5g}\\
\left[\frac{\delta}{\delta \tau} + \frac{\delta}{\delta
\lambda}\right] \Psi_{{\cal A}6} &=&0\, , \,
\left[\frac{\delta}{\delta {\widehat{P}}_0} +
e^{(-\frac{\lambda}{2}+\frac{\tau}{2})} {\widehat{P}}_0 \right]
\Psi_{{\cal A}6} = 0\, , \, \left[\frac{\delta}{\delta Q_0} +
e^{(-\frac{\lambda}{2}+\frac{\tau}{2})}
{\widehat{Q}}_0\right]\Psi_{{\cal A}6}= 0\label{e6g}\\
\left[\frac{\delta}{\delta \tau} - \frac{\delta}{\delta
\lambda}\right] \Psi_{{\cal A}7}&=& 0 \, , \,
\left[\frac{\delta}{\delta {\widehat{P}}_0} -
e^{(-\frac{\lambda}{2}+\frac{\tau}{2})} {\widehat{P}}_0 \right]
\Psi_{{\cal A}7} = 0 \, , \, \left[\frac{\delta}{\delta Q_0} +
e^{(-\frac{\lambda}{2}+\frac{\tau}{2})}
{\widehat{Q}}_0\right]\Psi_{{\cal A}7} = 0\label{e7g}\\
\left[\frac{\delta}{\delta \tau} - \frac{\delta}{\delta
\lambda}\right] \Psi_{{\cal A}8} &=&0 \, , \,
\left[\frac{\delta}{\delta {\widehat{P}}_0} +
e^{(-\frac{\lambda}{2}+\frac{\tau}{2})} {\widehat{P}}_0 \right]
\Psi_{{\cal A}8} = 0 \, , \, \left[\frac{\delta}{\delta Q_0} -
e^{(-\frac{\lambda}{2}+\frac{\tau}{2})}
{\widehat{Q}}_0\right]\Psi_{{\cal A}8}= 0\label{e8g}\,
\end{eqnarray}
It is straightforward to see that only Eqs. (\ref{e1g}),
(\ref{e2g}), (\ref{e5g}), and (\ref{e6g}) can be consistently
solved. Therefore, the physical state reads
\begin{eqnarray}
\Psi_{{\cal A}1} &=& \Psi_{{\cal A}10}\exp[m(\lambda - \tau)]
\exp\left[-e^{(-\frac{\lambda}{2}+\frac{\tau}{2})}
\left(\frac{{\widehat{P}}^2_0 -
{\widehat{Q}}^2_0}{2}\right)\right] \, , \\
\Psi_{{\cal A}2} &=& \Psi_{{\cal A}20}\exp[m(\lambda - \tau)]
\exp\left[-e^{(-\frac{\lambda}{2}+\frac{\tau}{2})}
\left(\frac{-{\widehat{P}}^2_0 + {\widehat{Q}}^2_0}{2}\right)\right]\, , \\
\Psi_{{\cal A}3} &=& 0 \, , \\
\Psi_{{\cal A}4} &=& 0\, ,\\
\Psi_{{\cal A}5} &=& \Psi_{{\cal A}50}\exp[m(\lambda - \tau)]
\exp\left[e^{(-\frac{\lambda}{2}+\frac{\tau}{2})}
\left(\frac{{\widehat{P}}^2_0 +
{\widehat{Q}}^2_0}{2}\right)\right] \, , \\
\Psi_{{\cal A}6} &=& \Psi_{{\cal A}60}\exp[m(\lambda - \tau)]
\exp\left[-e^{(-\frac{\lambda}{2}+\frac{\tau}{2})}
\left(\frac{{\widehat{P}}^2_0 + {\widehat{Q}}^2_0}{2}\right)\right] \, , \\
\Psi_{{\cal A}7} &=& 0 \, , \\
\Psi_{{\cal A}8} &=& 0\, ,
\end{eqnarray}
or equivalently
\begin{equation}
\Psi_{{\cal A}}=\exp[m(\lambda - \tau)]\left( \begin{array}{c}
\Psi_{{\cal A}10}\exp[-e^{(-\frac{\lambda}{2}+\frac{\tau}{2})}
(\frac{{\widehat{P}}^2_0 - {\widehat{Q}}^2_0}{2})]\\
\Psi_{{\cal A}20}\exp[-e^{(-\frac{\lambda}{2}+\frac{\tau}{2})}
(\frac{-{\widehat{P}}^2_0 + {\widehat{Q}}^2_0}{2})]\\
0 \\
0\\
\Psi_{{\cal A}50}\exp[e^{(-\frac{\lambda}{2}+\frac{\tau}{2})}
(\frac{{\widehat{P}}^2_0 + {\widehat{Q}}^2_0}{2})]\\
\Psi_{{\cal A}60}\exp[-e^{(-\frac{\lambda}{2}+\frac{\tau}{2})}
(\frac{{\widehat{P}}^2_0 + {\widehat{Q}}^2_0}{2})]\\
0\\
0\\
\end{array}\right) \label{sol2g}\, ,
\end{equation}
where $\Psi_{{\cal A}10}$, $\Psi_{{\cal A}20}$, $\Psi_{{\cal
A}50}$ and $\Psi_{{\cal A}60}$ are integration constants and $m$
is, as before, a separation constant. Fig.~\ref{solgenplot} shows
explicitly the behavior of the solution (\ref{sol2g}).

%***************************************************************
\section{Final remarks and conclusions}
\label{con}

The common concern about the use of this formalism is whether the
final result of the quantization (in our case, the state function
of the universe) depends on the choice of a particular foliation.
To clarify this issue in the present case let us consider another
choice of the spacetime foliation, i.e., a new time coordinate $t$
defined by \begin{equation} dt = e^{-\frac{1}{4}(\lambda + 3\tau)}
d\tau \ . \label{ttran} \end{equation} Then, from the general line
element (\ref{gle}) we obtain \begin{equation} ds^2 =  d t^2 -
e^{2\Lambda} d\chi^2 -  e^F \left(e^P d\sigma ^2 + e^{-P}
d\delta^2\right)\ , \label{gle1} \end{equation} where
$\Lambda=\Lambda(t,\chi)$ and $F=F(t,\chi)$. For the sake of
simplicity we are considering here the polarized case only
$(Q=0)$. In this particular parametrization the lapse function
becomes a constant. Consequently, the problem of the {\em frozen
time} of the canonical quantization cannot be solved in the
framework of our present analysis.

It is dangerous to draw conclusions from some models,
minisuperspace or even midisuperspace ones, to full quantum
gravity. One should try to avoid common practice, which consists
of solving a time problem for a model way down in the hierarchy,
and jumping to the conclusion that the time problems of quantum
gravity are removed by the same treatment.

On the other hand, it is important to emphasize that the physical
interpretation of the wave function of the universe
$\vert\Psi\rangle$ presents certain difficulties.  A genuine wave
function must be related to observable quantities and this implies
that $\vert\Psi\rangle$ must yield a probability density. However,
this is not true in this case, since the wave function of the
universe is not normalizable. Moreover, if we require that
$\vert\Psi\rangle$ yields a probability density for the
3--geometry, which, as it is usual in quantum field theory, must
have a specific value at a given time, this would imply a
violation of the Hamiltonian constraint \cite{kucryan}. These
difficulties in the interpretation of the state function of the
universe are the price one has to pay for the use of the canonical
quantization procedure, and its inherent a preferred foliation,
i.e., the isolation of a specific {\em absolute time} parameter
against which the evolution of the system should be defined. An
alternative procedure like the Dirac quantization, based on
functional integrals, does not require to single out the time
variable and could lead to a quantum system with less
interpretation difficulties \cite{guvryan}. Nevertheless, even
this other Dirac approach does not solve the time arbitrariness
problem.

In this work we have investigated the quantization of Gowdy $T^3$
cosmological models in the context of $N=1$ supergravity. The
quantum constraints, resulting from the canonical quantization
formalism, are explicitly analyzed and solved. In this way, we
find the state function of the universe for the polarized and
unpolarized Gowdy $T^3$ models. This represents a proof of the
existence of physical states in the $(N=1)$ supersymmetric simple
midisuperspace, corresponding to Gowdy cosmologies. This result
contrasts drastically with analogous investigations in
minisuperspace (Bianchi--like) models, where no physical states
exist, a result that  sometimes is assumed as a sufficient proof
to dismiss $N=1$ supergravity. We have adopted a less radical
position in this work and dismiss as non--physical only the
homogeneous minisuperspace models. The existence of physical
states in midisuperspace models confirms this conclusion and
indicates that $N=1$ supergravity is a valuable theory which
should be investigated further. In this context we have also
obtained an interesting result showing that, in the Gowdy $T^3$
midisuperspace model analyzed in this work, the state function of
the universe, representing nontrivial physical states is
completely free of anomalies.

On the other hand, there exists a belief that the second
quantization solves the problem of time in quantum theory of a
relativistic particle. The second quantization approach to quantum
field theory is based on the construction of a Fock space, i.e.,
one takes a one--particle Hilbert space ${\cal F}_{(1)}$. From the
direct product of the one--particle states the states which span
the N--particle sector ${\cal F}_{(N)}$ are constructed. The Fock
space ${\cal F}$ is then the direct sum of all such sectors, i.e.,
${\cal F} = {\cal F}_{(0)} \oplus {\cal F}_{(1)} \oplus {\cal
F}_{(2)} \oplus \cdots$, where ${\cal F}_{(0)}$ is spanned by the
vacuum state. It is clear that the Fock space ${\cal F}$ can be
defined only if the one--particle state ${\cal F}_{(1)}$ is a
Hilbert space. This brings us to the Hilbert space problem for a
relativistic particle. The absence of a privileged one--particle
Hilbert space structure is source of ambiguities in constructing a
unique quantum field theory on a dynamical background
\cite{Kuchar}.

In full, the second quantization merely shifts the problem of the
arbitrariness of time to a different level without really solving
it. Consequently, our quantization approach \`a la Pilati
\cite{pi78} remains valid since the second quantization does not
represent a significant improvement to the quantization approach
regarding the time evolution problem.

A closer look to the second quantization approach reveals that it
does not really solve the problem of time evolution and its
formalism resists an operational interpretation, like the problems
presented by the indefinite inner product of the Klein--Gordon
interpretation, which are faced by suggesting that the solutions
of the Wheeler--DeWitt equation are to be turned to operators.
This is analogous to subjecting the relativistic particle, whose
state is described by the Klein--Gordon equation, to second
quantization.

In this work we have focused on the special case of $T^3$
cosmologies. The generalization of our results to include the case
of $S^1\times S^2$ Gowdy models seems to be straightforward. In
particular, we believe that the unified parametrization introduced
in \cite{procmike}, which contains both types of topologies, could
be useful to explore the supersymmetric Gowdy model in quite
general terms.

%*************************************************************
\begin{acknowledgments}

This research was supported by CONACyT Grants 48404--F and
47000-F, and by the collaboration Mexico--Germany, grants
CONACyT--DFG J110.491 and J110.492.

\end{acknowledgments}

%**************************************************************

\end{document}